\documentclass{article}

\usepackage{arxiv}

\usepackage{graphicx}
\usepackage{subcaption}
\usepackage{array,colortbl,multirow,multicol,booktabs,ctable}
\usepackage{afterpage}
\usepackage{float}
\usepackage[utf8]{inputenc} % allow utf-8 input
\usepackage[T1]{fontenc}    % use 8-bit T1 fonts
\usepackage{hyperref}       % hyperlinks
\usepackage{url}            % simple URL typesetting
\usepackage{booktabs}       % professional-quality tables
\usepackage{amsfonts}       % blackboard math symbols
\usepackage{nicefrac}       % compact symbols for 1/2, etc.
\usepackage{microtype}      % microtypography
\usepackage{graphicx}
\usepackage{natbib}
\usepackage{doi}
\usepackage{amsmath}

\title{GAL-MAD: Towards Explainable Anomaly Detection in Microservice Applications Using Graph Attention Networks}

\author{Lahiru Akmeemana \\
	Department of Computer Science and Engineering\\
	University of Moratuwa\\
	Sri Lanka\\
	\texttt{lahirua.18@cse.mrt.ac.lk} \\
	\And
	Chamodya Attanayake \\
	Department of Computer Science and Engineering\\
	University of Moratuwa\\
	Sri Lanka\\
	\texttt{chamodya.18@cse.mrt.ac.lk} \\
	\And
	Husni Faiz \\
	Department of Computer Science and Engineering\\
	University of Moratuwa\\
	Sri Lanka\\
	\texttt{husni.18@cse.mrt.ac.lk} \\
    \And
	Sandareka Wickramanayake \\
	Department of Computer Science and Engineering\\
	University of Moratuwa\\
	Sri Lanka\\
	\texttt{sandarekaw@cse.mrt.ac.lk} \\
}

\date{}

\renewcommand{\headeright}{A PREPRINT}
\renewcommand{\undertitle}{A PREPRINT}
\renewcommand{\shorttitle}{\textit{GAL-MAD} Explainable Anomaly Detection in Microservice Applications Using GAT}

\hypersetup{
pdftitle={GAL-MAD: Towards Explainable Anomaly Detection in Microservice Applications Using Graph Attention Networks},
pdfkeywords={Anomaly detection, Microservices, Multivariate data, Time series data, Graph
Attention, LSTM, Autoencoder, Explainability},
}

\begin{document}
\maketitle

\begin{abstract}
	The distributed and dynamic nature of microservices introduces complexities in ensuring system reliability, making anomaly detection crucial. Existing anomaly detection techniques often rely on statistical models or machine learning methods that struggle with high-dimensional and interdependent data. Available datasets have limitations in capturing network and performance metrics of microservices. This work introduces the RS-Anomic dataset generated using the open-source RobotShop microservice application. The dataset captures multivariate performance metrics and response times under normal and anomalous conditions, encompassing ten types of anomalies. We propose a novel anomaly detection model called Graph Attention and LSTM-based Microservice Anomaly Detection (GAL-MAD), leveraging Graph Attention and Long Short-Term Memory architectures to capture spatial and temporal dependencies in microservices. Using SHAP, we explore anomaly localization to enhance explainability. Experimental results demonstrate that GAL-MAD outperforms state-of-the-art models on the RS-Anomic dataset, achieving higher accuracy and recall across varying anomaly rates.     
\end{abstract}

% keywords can be removed

\section{Introduction}

The migration from monolithic to microservice architecture has become a prevalent trend in cloud applications due to benefits such as improved scalability, modularity, and autonomous components that simplify development. A few enterprise companies using microservice architecture include Amazon, Netflix, and Uber. A microservice application consists of loosely coupled, independently deployable services, with each microservice performing a specific functionality or a business objective. Typically the performance of microservices are guarded by Service Level Objectives (SLOs).

When a system experiences a fault or is being used unexpectedly, it can result in erroneous behaviour, also known as an anomaly. Detecting these anomalies is essential to ensure the entire microservice application functions correctly and efficiently. Several factors can contribute to anomalous behaviour, such as hardware faults, misconfigurations, and unexpected load. However, identifying and resolving these anomalies can be challenging due to the complexity of the microservices architecture, which involves multiple independent services working together, asynchronous communication, frequent changes, and complex dependencies. Even a minor change in one service can significantly impact others, making it difficult to isolate the root cause of the anomaly.

The existing datasets for anomaly detection in microservice applications primarily consist of system traces and logs, and most of them are not publicly available \cite{zhao2020multivariate, wu2020microrca, Kohyarnejadfard2022}. Further, many of these datasets contain only univariate time-series data, which lack sufficient detail and do not capture the interdependencies among services. As a result, they can lead to suboptimal performance in anomaly detection \cite{zhao2020multivariate}. Instead, a publicly available multivariate time-series dataset that includes microservice application performance data would significantly enhance anomaly detection in such applications.

Additionally, current approaches to anomaly detection in microservices primarily rely on statistical and machine learning techniques that focus on univariate or log-based data. While these methods can be effective in specific situations, they often overlook the complex interdependencies and high-dimensional characteristics of microservices. The accuracy and efficiency of these anomaly detection methods can vary significantly depending on the application domain and the type of data collected. Moreover, existing methods typically do not provide insights into the root causes of detected anomalies. Without a clear understanding of these root causes, system administrators face challenges in implementing timely and effective resolutions.

This research presents the \textit{RS-Anomic dataset}, which contains multivariate data from a microservice application and a framework for detecting anomalous behaviour of a microservice application and localizing the affected microservice. We utilized RobotShop \cite{Robotshop} as the microservice application for this dataset. RobotShop is a small-scale microservice application that aims to teach containerized application orchestration and monitoring methods. RS-Anomic provides performance metrics on memory, CPU utilization, file Input/Output, and network as a multivariate time series for each service. These metrics can be valuable in developing novel anomaly detection methods that explore inter-dependencies between services and their performance metrics.

Moreover, we introduce a new model called Graph Attention and LSTM-based Microservice Anomaly Detection (GAL-MAD) for detecting anomalies in microservices. This model is an encoder-decoder architecture trained using the performance metrics of the microservice system during normal operation. When provided with the performance metrics over a specific period, the model learns to reconstruct this input accurately. During inference, if the system's performance metrics indicate anomalous behaviour, the model exhibits a higher reconstruction loss, aiding in identifying such anomalies. The encoder component employs Graph Attention layers to capture the dependencies among cooperating microservices, and Long Short-Term Memory layers to recognize the temporal patterns in the operations of the microservice system. The decoder mirrors the structure of the encoder. To assess the performance of our model, we compare it with state-of-the-art anomaly detection models using the RS-Anomic dataset, highlighting the importance of capturing interdependencies within the microservices domain. Additionally, we utilize SHapley Additive exPlanations \cite{shap} for root cause analysis, reinforcing the model's predictions and enhancing its practical utility.

\section{Related Work}

This section presents the existing work related to our research under three categories: datasets for anomaly detection, anomaly detection methods, and root cause analysis.

\subsection{Datasets for Anomaly Detection}

Performance and response time features play a critical role in detecting and localizing anomalies in microservices. Several large-scale datasets are available for anomaly detection using microservice trace logs \cite{illinoisdatabankIDB-6738796, meta-microservices, lee_2024_13947828}. These trace logs enable the identification of slow or faulty services by capturing end-to-end service call behaviors. As traces provide unified measurements of response time, including latency and overall performance, they are effective for service-level anomaly detection. However, their utility for root cause analysis is often limited to pinpointing the anomalous service or service call, without offering deeper insights into system-wide issues.

To gain a more holistic understanding of system behavior, multivariate datasets capturing system-level performance metrics are essential. Response time remains a key component of latency-related Service Level Objectives (SLOs) and is central to many anomaly detection techniques. For instance, the NAB dataset collection from Numenta \cite{lavin2015evaluating} includes server metrics collected via Amazon CloudWatch but lacks response time metrics. In contrast, Luo et al. \cite{luo2021characterizing} introduced a more comprehensive dataset that combines system resource utilization with response time data. However, it does not capture network resource usage, which is crucial for detecting subtle or fine-grained network anomalies. Therefore, this research introduces a new multivariate dataset capturing performance metrics of a microservice network to improve anomaly detection effectiveness.

\subsection{Anomaly Detection}

Anomaly detection focuses on identifying observations that significantly diverge from normal patterns, a task central to fraud detection, cybersecurity, and healthcare monitoring. Traditional approaches rooted in statistical tests and shallow Machine Learning (ML) classifiers often struggle with high dimensional or unstructured data and the severe class imbalance inherent to anomaly detection tasks. Deep learning has revolutionized anomaly detection in recent years by learning hierarchical representations that capture complex data distributions, yielding more accurate and scalable solutions \cite{10.1145/3439950}. A prevalent paradigm employs pre-trained neural networks, such as CNNS or RNNS, as feature extractors, whose embeddings are subsequently fed into ML classifiers like One-Class Support Vector Machines to detect outliers \cite{8721681}. Reconstruction based models, including autoencoders and variational autoencoders, train solely on normal data and flag inputs with high reconstruction error as anomalies, \cite{chen2017outlier,Kitsune}. Prediction based techniques leverage forecasting networks often LSTM autoencoders that predict future time steps, denoting large prediction deviations as anomalies \cite{li2019mad,ergen2019unsupervised}. Generative adversarial frameworks learn the distribution of normal data via adversarial training, using both generator reconstruction losses and discriminator confidences for anomaly scoring \cite{li2019mad,schlegl2019f}. End-to-end one-class neural methods such as Deep SVDD and its semi-supervised extension DeepSAD unify representation learning and scoring by optimizing hypersphere or ranking objectives to tightly enclose normal instances in latent space \cite{chi2024deep}. This research explores the approaches that learn feature representations of normality for anomaly detection, specifically focusing on those based on graph data.

The Graph Deviation Network (GDN) employs a graph neural network architecture to learn inter-variable relationships in multivariate time series via attention-based forecasting, using a structure learned from cosine similarity to compute deviation scores for anomaly detection tasks with both high accuracy and explainability \cite{deng2021graph}. The Multivariate Anomaly Detection with Generative Adversarial Network (MAD-GAN) \cite{li2019mad} integrates LSTM-based generator and discriminator networks within a GAN framework to jointly capture temporal dependencies and distributional interactions among variables, producing a novel DR-score that combines reconstruction error and discriminator confidence to flag anomalies. Kitsune\cite{Kitsune}, an unsupervised network intrusion detection system, utilizes the KitNET autoencoder ensemble, trained incrementally on real-time network traffic features extracted by the AfterImage framework, to reconstruct normal traffic patterns, identifying deviations through elevated reconstruction errors and ensemble based thresholding, all with minimal computational overhead suitable for edge devices. Chen et al. \cite{chen2017outlier} introduced autoencoder ensembles for unsupervised outlier detection, demonstrating that randomly varying the connectivity architectures of autoencoders and employing adaptive sampling enhances robustness to noise and improves detection performance across benchmark datasets. Underpinning these graph-based models, Graph Attention Networks (GAT) extend traditional graph neural networks by incorporating masked self-attention mechanisms that assign learnable importance weights to neighboring nodes, enabling inductive learning on graphs with complex, heterogeneous structures without prior knowledge of the graph topology \cite{velickovic2017graph}.

A growing body of work addresses anomaly detection in microservice systems by leveraging unstructured logs, structured performance metrics, or a combination of both. Log-based methods evolved from statistical and clustering techniques (e.g., PCA, invariant mining) to deep sequence and graph models such as DeepLog \cite{du2017deeplog} and DeepTraLog\cite{zhang2022deeptralog}, which capture temporal and structural patterns in service calls. Metric based approaches rely on supervised and unsupervised learning over CPU, memory, I/O, and response-time series, ranging from simple MLP classifiers\cite{nobre2023anomaly} to NLP-based span analysis and distributed trace profiling. Recent hybrid methods integrate logs, traces, and metrics to exploit complementary views of system behavior\cite{li2024multi}. This research explores anomaly detection using performance metrics.

Performance metrics based approaches monitor resource and latency indicators such as CPU, memory, disk I/O, network, and response time, and apply ML to detect deviations. Supervised models like simple Multi-Layer Perceptrons trained on labelled metric traces can accurately identify resource bottlenecks and failures. Alternatively, methods leveraging natural language processing on trace spans treat latency values as sequences, applying NLP techniques to spot performance regressions without prior system knowledge.

\subsection{Root Cause Analysis}

Modern deep anomaly detectors for microservices not only flag abnormal behaviors but increasingly aim to pinpoint why they occur, leveraging causal inference, graph reasoning, and multi-source fusion. Causal-discovery approaches such as CausalRCA \cite{xin2023causalrca} and RUN \cite{lin2024root} employ gradient-based or neural Granger methods to learn directed graphs from metrics and time series, then infer fine-grained metric-level causes within faulty services. Graph-pruning and reinforcement-learning-based RCA (TraceDiag\cite{ding2023tracediag}) achieve interpretable, end-to-end localization on large-scale systems by learning to trim irrelevant dependencies before causal analysis. Although these methods embed some interpretability (e.g., attention weights, pruned graphs), only a few works explicitly adopt post-hoc XAI tools for RCA in microservices, e.g., yRCA \cite{soldani2023yrca}. This paper explores the potential of anomaly localization through explainability using existing post-hoc explainability techniques such as SHapley Additive exPlanations (SHAP)\cite{lundberg2017unified}.

\section{Methodology}

\subsection{RS-Anomic Dataset}

The RS-Anomic dataset comprises performance metrics for 12 services, each with 19 performance metrics and a variable number of response time metrics per service, as detailed in Table \ref{table:feature_details}. Furthermore, the anomaly data in the RS-Anomic dataset covers ten anomalous behaviours that may occur in microservice applications. RS-Anomic contains 100464 normal and 14112 anomalous instances. Each microservice communicates with a different number of microservices, and the response times for each communication link are recorded in the dataset. The dataset and data loading scripts are available at \url{https://github.com/ms-anomaly/rs-anomic}. Normal data and anomalous data are contained in two zip files. Each zip file contains 2 folders named \textit{cAdvisor} and \textit{response\_times} containing performance metrics and response time data respectively. Response time and performance data should be concatenated to obtain the complete dataset. In the case of anomalous data, \textit{cAdvisor} and \textit{response\_times} folders are further divided into each anomaly type, to obtain the complete anomalous data, response times of each anomaly should be concatenated with the corresponding performance metrics.

\subsubsection{Testbed and Data Collection Environment}

\begin{figure}
    \centering
    \includegraphics[width=0.6\linewidth]{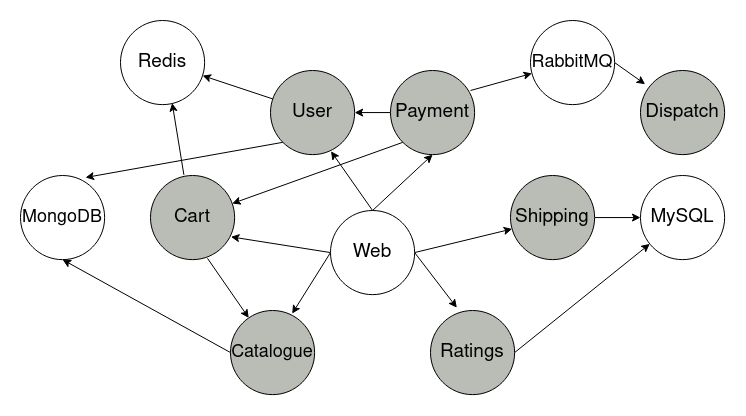}
    \captionsetup{font=small}
    \caption{\fontsize{10pt}{11pt}\selectfont{\itshape{RobotShop microservice architechture}}}
    \label{fig:architecture}
\end{figure}

The RS-Anomic dataset was created using RobotShop \cite{Robotshop}, a microservice application that implements containerized orchestration and monitoring techniques. RobotShop is an open-source project with an e-commerce web application built using a microservices architecture. All services are deployed as Docker containers on a single server, communicating through a Docker bridge network. The application comprises 12 services, with their dependencies illustrated in Figure \ref{fig:architecture}. This application was used in \cite{Harlicaj2021} to evaluate their model. The testbed was run on a server with a 40-core CPU and 64 GB Memory running Ubuntu 20.04.5.
We used Prometheus \cite{208870} for data acquisition. The Prometheus server was configured to poll data from the microservices every 5 seconds. We instrumented the response times in microservice calls using Prometheus client libraries. The instrumented services are highlighted in Figure \ref{fig:architecture}. Using Prometheus client libraries response time metrics were exposed from the microservices over HTTP endpoints. The cumulative summation of response time and the number of service calls that occurred between the data polling interval for each communication link was recorded.
Container Advisor (cAdvisor) \cite{cAdvisor} was utilized to gather runtime metrics. This tool offers measurements related to resource usage, such as CPU, memory, disk Input/Output (I/O), and network I/O of Docker containers. Additionally, data from cAdvisor was also polled from the Prometheus server. Data collection spanned over six days under normal conditions. For each type of anomalous behaviour, data was collected over 90 minutes by injecting faults into specific services to simulate anomalies. Table \ref{table:feature_details} shows the features available in the RS-Anomic dataset.
To ensure RS-Anomic covers an actual e-commerce application behaviour, we simulated a time-varying load with more users in peak hours and fewer in off-peak hours. We used Locust \cite{locust}, an open-source load testing tool, for load generation. The load generation can be done by creating test scenarios that simulate user behaviour and defining the number of virtual users and requests per second that should be generated during the test.

\begin{table*}[ht]
\centering
\begin{tabularx}{\textwidth}{|X|X|}
    \hline
\bfseries Feature                                               & \bfseries Description                     \\ \hline
container\_memory\_rss                                & Resident set size of a container in bytes at the polled time           \\ \hline
container\_memory\_usage\_bytes                       & Total memory usage in bytes at the polled time                         \\ \hline
container\_memory\_failures\_total                    & Cumulative count of memory allocation failures                 \\ \hline
container\_memory\_working\_set\_bytes                & Current working set size in bytes at the polled time                   \\ \hline
container\_memory\_failcnt                            & Cumulative count of memory usage exceeds limit                 \\ \hline
container\_cpu\_usage\_seconds\_total                 & Cumulative CPU usage seconds in total               \\ \hline
container\_cpu\_user\_seconds\_total                  & Cumulative user cpu time consumed                   \\ \hline
container\_cpu\_system\_seconds\_total                & Cumulative system cpu time consumed                 \\ \hline
container\_network\_receive\_bytes\_total             & Cumulative number of bytes received                                      \\ \hline
container\_network\_receive\_errors\_total            & Cumulative count of errors encountered while receiving         \\ \hline
container\_network\_receive\_packets\_dr-{\newline}opped\_total  & Cumulative count of packets dropped while receiving            \\ \hline
container\_network\_receive\_packets\_total           & Cumulative count of packets received                           \\ \hline
container\_network\_transmit\_bytes\_total            & Cumulative number of bytes transmitted                                   \\ \hline
container\_network\_transmit\_errors\_total           & Cumulative count of errors encountered while transmitting      \\ \hline
container\_network\_transmit\_packets\_d-{\newline}ropped\_total & Cumulative count of packets dropped while transmitting         \\ \hline
container\_network\_transmit\_packets\_t-{\newline}otal          & Cumulative count of packets transmitted                        \\ \hline
container\_fs\_usage\_bytes                           & Bytes used by a container on the file systems at the polled time       \\ \hline
container\_fs\_io\_time\_seconds\_total               & Cumulative time spend on file I/O                    \\ \hline
container\_fs\_write\_seconds\_total                  & Cumulative time spent on file writes                \\ \hline
Response times*                                       & Cumulative Response times to communicating services \\
    \hline
\end{tabularx}
%\captionsetup{font=small}
\caption{\fontsize{10pt}{11pt}\selectfont{\itshape{RS-Anomic feature descriptions. *Number of response times features may vary for each service based on the microservice architecture}}}
\label{table:feature_details}
\end{table*}

\subsubsection{Anomalous Behaviours Captured in RS-Anomic}

\begin{itemize}
 \item \textbf{Service Down:} When a microservice is not working and cannot respond to requests, it is referred to as a service down. Network problems, hardware malfunctions, or software glitches can cause service down. To simulate this issue, we manually terminated the service and observed how the remaining services continued functioning.

\item \textbf{High Concurrent User Load:} When a large unexpected number of users try to access the application simultaneously, it can cause slow response times, timeouts, or service failures. To simulate a high concurrent user load, we used our load generation script to simulate 1500 concurrent users accessing the application.

\item \textbf{High CPU usage:} A microservice with a poorly optimized algorithm, inefficient code, or increased load can cause the entire application to underperform, resulting in high CPU usage. To simulate high CPU usage, we ran 100 parallel threads that calculate the 1000000th Fibonacci number.

\item \textbf{High File I/O:} This anomaly occurs when a microservice performs excessive file I/O operations leading to performance issues, such as slow response times and high CPU usage. To simulate high file I/O, we used a thread to continuously read from and write to a file.

\item \textbf{Memory Leak:} This fault occurs when a microservice fails to release memory that is no longer needed. Over time, this can increase memory usage and cause the service to crash or become unresponsive. Stress-ng \cite{stress-ng} is used to continuously allocate memory without deallocating, which will simulate a memory leak.

\item \textbf{Packet Loss:} This anomaly occurs when packets of data sent between the microservice and other systems are lost due to network issues which result in slow response times, errors, or incomplete transactions. We use Traffic Control(tc) \cite{trafficcontrol} to simulate network conditions where most packets are lost.

\item \textbf{Response Time Delay:} Response time delay occurs when a microservice takes longer than usual to respond to requests due to various factors, including high CPU usage, increased user load, network latency, or software bugs. We simulate this behaviour by adding a delay time to API service calls.

\item \textbf{Out-of-Order Packets:} Out-of-order packets are a common occurrence in microservices due to the distributed nature of the system and the use of asynchronous communication methods.

\item \textbf{Low Bandwidth:} A significant deviation from the expected bandwidth usage can be caused by inefficient communication protocols, lack of load balancing, network hardware limitations, or improper network configurations.

\item \textbf{High Latency:} Microservices rely heavily on network communication to interact with each other, and any increase in latency can cause delays in the response time of the microservices, leading to a degraded user experience.  
\end{itemize}

We observed that network packet loss, out-of-order packets, and high latency anomalies do not clearly distinguish between normal and anomalous data. Hence, we increased the strength of these anomalies. For example, we increased the latency from 200ms to 1000ms as a 200ms latency anomaly is more challenging to distinguish from normal data points.

\subsection{Graph Attention and LSTM-based
Microservice Anomaly Detection (GAL-MAD) Model}

Reproducing anomalies for various microservice applications is often impractical. Hence, the recent work has leveraged unsupervised approaches to develop anomaly detection models for microservices. Following them, we develop an unsupervised framework that uses the performance metrics of normally operating services for model training. Our model, Graph Attention and LSTM-based Microservice Anomaly Detection (GAL-MAD), employs an autoencoder architecture to learn a compact representation of normal behaviour and identify deviations as anomalies at test time.

\begin{figure*}[!t]
    \includegraphics[width=\linewidth]{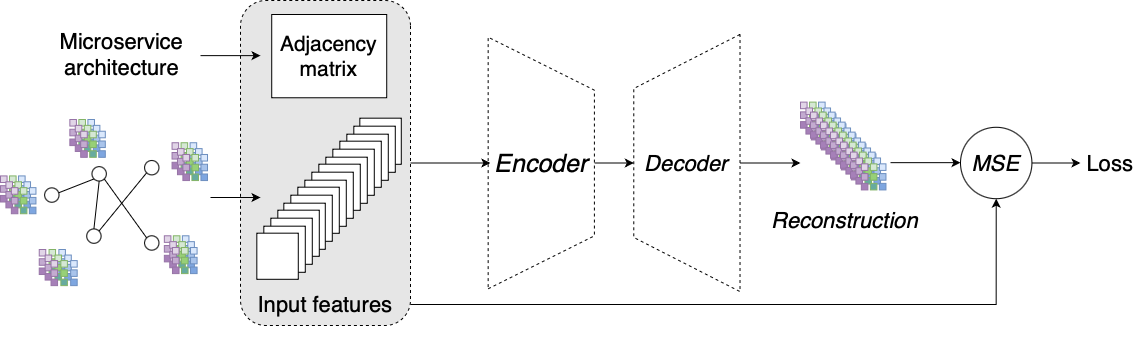}
    \centering
    \caption{GAL-MAD Model architecture}
    \label{fig:final_model}
\end{figure*}

We consider a microservice system comprising \( n \) services, each characterized by \( k \) performance metrics. Over \( t \) time steps, we collect data representing the system's behavior. GAL-MAD is structured as an autoencoder with an encoder-decoder architecture. The encoder utilizes two Graph Attention Network (GAT) layers \cite{velickovic2017graph} followed by a bidirectional Long Short-Term Memory (Bi-LSTM) layer to capture spatial and temporal dependencies among services. The decoder mirrors the encoder's architecture with a Bi-LSTM layer followed by two GAT layers applied in reverse order. GAL-MAD is trained on normal data to minimize the reconstruction loss using Mean Squared Error (MSE). The overall architecture of GAL-MAD is shown in Figure. \ref{fig:final_model}.

Let $\mathbf{X}^{(i)} \in \mathbb{R}^{n \times k} $ denote the feature matrix at time step $i$, where each row corresponds to a microservice's features and $\mathbf{A} \in \{0,1\}^{n \times n} $ represent the adjacency matrix encoding the static topology of microservice interactions. We stack the feature matrices over $t$ time steps to form a 3D tensor.

\begin{equation}
\mathcal{X} = \left[ \mathbf{X}^{(1)}, \mathbf{X}^{(2)}, \ldots, \mathbf{X}^{(t)} \right] \in \mathbb{R}^{t \times n \times k}
\end{equation}

The encoder processes the input tensor $\mathcal{X}$ through two sequential GAT layers to learn latent representations, followed by a Bi-LSTM layer to capture temporal dependencies.

For each time step $ i \in \{1, \ldots, t\} $, the first GAT layer computes the new node embeddings as follows.
\begin{equation}
\mathbf{H}^{(1)}_i = \text{GAT}_1\left( \mathbf{X}^{(i)}, \mathbf{A} \right) \in \mathbb{R}^{n \times d_1}
\end{equation}
, where $\text{GAT}_1$ denotes the first GAT layer and $d_1$ is the output feature dimension of the first GAT layer. After stacking over $t$ time steps, the output of the first GAT layer, $\mathcal{H}^{(1)}$, is

\begin{equation}
\mathcal{H}^{(1)} = \left[ \mathbf{H}^{(1)}_1, \mathbf{H}^{(1)}_2, \ldots, \mathbf{H}^{(1)}_t \right] \in \mathbb{R}^{t \times n \times d_1}
\end{equation}

Similarly, the second GAT layer, $\text{GAT}_2$, processes $\mathcal{H}^{(1)}$ to produce $\mathcal{H}^{(2)}$

\begin{equation}
\mathcal{H}^{(2)} = \left[ \mathbf{H}^{(2)}_1, \mathbf{H}^{(2)}_2, \ldots, \mathbf{H}^{(2)}_t \right] \in \mathbb{R}^{t \times n \times d_2}
\end{equation}
, where \( d_2 \) is the output feature dimension of the second GAT layer.

To capture temporal dependencies, we reshape \( \mathcal{H}^{(2)} \) to a sequence suitable for the Bi-LSTM layer as follows.
\begin{equation}
\mathcal{H}^{(2)}_{\text{reshaped}} \in \mathbb{R}^{n \times t \times d_2}
\end{equation}

\begin{figure*}[!t]
    \includegraphics[width=\linewidth]{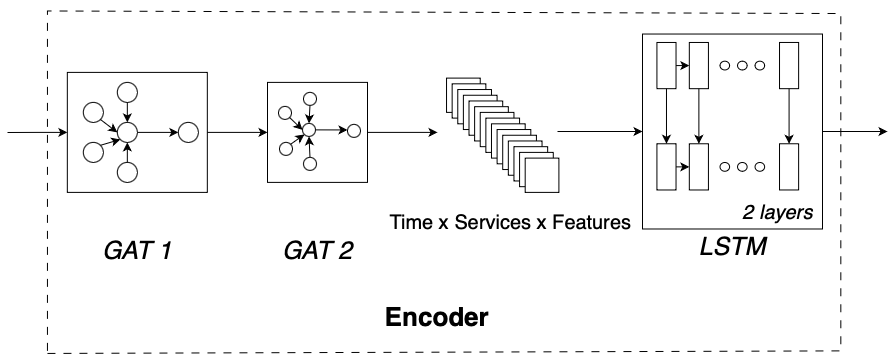}
    \centering
    \caption{Encoder component of the GAL-MAD model}
    \label{fig:encoder_component}
\end{figure*}

The Bi-LSTM processes each service's sequence over time to produce $\mathbf{z}_j$
\begin{equation}
\mathbf{z}_j = \text{BiLSTM}\left( \mathcal{H}^{(2)}_{\text{reshaped}}[j, :, :] \right) \in \mathbb{R}^{d_z}
\end{equation}
, for \( j = 1, \ldots, n \), where \( d_z \) is the dimension of the embedding vector for each service. After stacking the embeddings, we get the final embedding of the encoder $\mathbf{Z} = \left[ \mathbf{z}_1, \mathbf{z}_2, \ldots, \mathbf{z}_n \right]^\top \in \mathbb{R}^{n \times d_z}$. Figure \ref{fig:encoder_component} elaborates on the encoder of the GAL-MAD.

The GAL-MAD decoder reconstructs the input by mirroring the encoder's architecture, which comprises a one-to-many LSTM followed by two GAT layers. 

The GAL-MAD model is trained on normal data, using MSE. Let the reconstruction error of GAL-MAD be $\mathcal{L}$. During inference, a higher MSE means the input is anomalous. The Sigmoid function is applied to MSE during inference to derive the final prediction as follows:

\begin{equation}
   y = Sigmoid(\mathcal{L}-c) 
\end{equation}
, where $c$ is the upper bound of the reconstruction loss observed for normal data. In our experiments, we set $c=2.0$. 

\subsubsection{Implementation Details}

The model was trained using the Adam optimiser with a batch size of 360, a learning rate of 0.001, and a decay of 0.5 per epoch. Each LSTM uses a sequence length 24, translating to 2-minute time windows polled every 5 seconds. Further, we set $d_z$ to 1. The GAL-MAD predicts anomalies in 80-second windows.

\subsection{Anomalous Service Localization}
\label{anomaly_service_locatiozation}

To enhance the interpretability of anomaly detection in microservice architectures, we employ SHapley Additive exPlanations (SHAP) \cite{shap} to localize anomalous services and identify contributing features. Upon detecting an anomaly, SHAP values are computed for each time step within the observation period, quantifying the contribution of each feature to the model's output. Time steps identified as anomalous by the detection model are selected for further analysis. The SHAP values corresponding to these anomalous time steps are aggregated along the temporal axis to obtain a consolidated view of feature contributions. To emphasize critical features, specific SHAP values are weighted; in particular, the SHAP values for the $response time$ and $container\_fs\_usage\_bytes$ features are multiplied by a factor of four. Subsequently, the absolute SHAP values are summed across features for each service to determine the overall contribution, with the service corresponding to the maximum sum identified as the most likely source of the anomaly. Within this anomalous service, the feature with the highest absolute SHAP value is determined, pinpointing the specific metric most responsible for the detected anomaly. This methodology facilitates the pinpointing of both the anomalous service and the specific feature contributing to the detected anomaly, thereby enhancing the explainability and actionable insights derived from the anomaly detection model.

\section{Experimental Study} 

\subsection{Experiments on RS-Anomic Dataset}
We evaluated the RS-Anomic dataset using state-of-the-art anomaly detection models, GDN \cite{deng2021graph}, MAD-GAN \cite{li2019mad}, Kitsune \cite{Kitsune} and a Transformer-based reconstruction model \cite{vaswani2017attention}. GDN uses cosine similarity to create a graph structure, predicts future values using graph attention-based forecasting, and calculates the graph deviation score. MAD-GAN is designed to detect anomalies in time series data and uses an LSTM-RNN discriminator and a reconstruction based method. Kitsune is an unsupervised Network Intrusion Detection System that can detect local network attacks without supervision. We categorized anomalies as the positive class and compared the models using Accuracy (A), Recall (R), and Specificity (S) metrics. The normal data was split into an 80:20 ratio for training and testing. Only normal data was used to train models based on respective papers. We created three testing scenarios with normal:anomalous behaviour ratios of 95:5, 90:10, and 60:40. For each testing scenario, the test data was mixed with anomalous data, according to the respective ratios. Anomalous samples were balanced across different anomaly types.
The difference in the number of response time features for each microservice, due to varying numbers of links between microservices, was unified by summing all recorded response times for each service to create one response time feature. If a service experienced a network delay in any communication link, this information was preserved by summing the response times. In addition to response times, we introduced a moving average of response times for 5 minute and 30 minute windows. The input dimensions for all the models were 264 for each time step assembled by concatenating 22 features, including moving averages for 12 services. Moreover, standard scaling to scale our data based on normal data distribution. We considered a window size of 24 data points. We trained all models for 20 epochs with Adam optimizer and learning rates 0.001, 0.0005, 0.1 and 0.05 for GDN, MAD-GANs, Kitsune and Transformer, respectively.
The results for GDN, MAD-GAN, Vanilla transformer, Kitsune and GAL-MAD models on the RS-Anomic dataset are displayed in Table \ref{table:dataset_results_transposed}. According to the results, especially the recall value, state-of-the-art anomaly detection models are not very successful in detecting anomalies in microservices. This indicates that RS-Anomic is a challenging dataset, and further research is necessary to develop new Machine Learning models for microservice anomaly detection. The suboptimal performance of current models may be attributed to their inability to account for the graph-like structure inherent in the RS-Anomic dataset. This structure, derived from the architecture of microservice applications, is critical for identifying dependencies among input features.

While the RS-Anomic dataset enables anomaly detection in microservice applications and allows for the development of an anomaly classification model, it is essential to note that our application was deployed on a single server. Therefore, the RS-Anomic dataset does not capture the impacts of a distributed deployment. Most microservice applications operate in distributed environments with auto-scaling, which introduces additional complexities not reflected in the RS-Anomic dataset.

\begin{table*}[h!tb]
\centering
\setlength\tabcolsep{6pt}
\begin{tabular}{|l|c|c|c|c|c|c|}
\hline
\bfseries{Ratio} & \bfseries{Metric} & \bfseries{GDN} & \bfseries{MAD-GAN} & \bfseries{Kitsune} & \bfseries{Transformer} & \bfseries{GAL-MAD} \\ \hline
\multirow{3}{*}{95:5} 
  & A   & \textbf{0.9961} & 0.6073 & 0.9757 & 0.9810 & 0.9838 \\ \cline{2-7}
  & R   & 0.8093 & 0.8627 & 0.9006 & 0.8252 & \textbf{0.9884} \\ \cline{2-7}
  & S   & 0.9645 & 0.5920 & 0.9804 & \textbf{0.9905} & 0.9559 \\ \hline
\multirow{3}{*}{90:10} 
  & A   & 0.9409 & 0.5669 & 0.8984 & 0.9678 & \textbf{0.9788} \\ \cline{2-7}
  & R   & 0.7684 & 0.6893 & 0.5529 & 0.7994 & \textbf{0.9228} \\ \cline{2-7}
  & S   & 0.9620 & 0.5521 & 0.9412 & \textbf{0.9885} & 0.9855 \\ \hline
\multirow{3}{*}{60:40} 
  & A   & 0.8361 & 0.5225 & 0.5921 & 0.8712 & \textbf{0.9500} \\ \cline{2-7}
  & R   & 0.7766 & 0.5238 & 0.4142 & 0.7491 & \textbf{0.8938} \\ \cline{2-7}
  & S   & 0.8741 & 0.5258 & 0.7077 & 0.9506 & \textbf{0.9859} \\ \hline
\end{tabular}
\caption{\fontsize{10pt}{11pt}\selectfont{\itshape{Performance of GAL-MAD and state-of-the-art anomaly detection models on the RS-Anomic dataset across varying anomaly-to-normal data ratios (95:5, 90:10, 60:40), measured in terms of Accuracy (A), Recall (R), and Specificity (S).}}}
\label{table:dataset_results_transposed}
\end{table*}

\begin{table*}[!htb]
\centering
\setlength\tabcolsep{6pt}
{\begin{tabular}{|l|l|l|}
\hline
\bfseries{Anomaly/Normal}                                                                   & \bfseries{Model without response times} & \bfseries{Model with response times} \\ \hline
Normal                                                                           & 1.974                        & 1.930                     \\
\hline
Service Down                                                                     & 17554.017                    & 40366.418                 \\
\hline
High CPU usage                                                                   & 353297.431                   & 338129.254                \\
\hline
\begin{tabular}[c]{@{}l@{}}High Concurrent User\\ Load (1500 users)\end{tabular} & 30781.602                    & 31068.059                 \\
\hline
High File I/O                                                                    & 278496.914                   & 264153.407                \\
\hline
\begin{tabular}[c]{@{}l@{}}Memory Leak (upto\\ 300mb)\end{tabular}               & 242.596                      & 224.751                   \\
\hline
Packet Loss(50\%)                                                                & 2.045                        & 157.540                   \\
\hline
Packet Loss(80\%)                                                                & 11.597                       & 23703.363                 \\
\hline
\begin{tabular}[c]{@{}l@{}}Response Time\\ Delay($\sim$400ms)\end{tabular}       & 1.854                        & 27714.654                 \\
\hline
\begin{tabular}[c]{@{}l@{}}Out of Order\\ Packets(25\%)\end{tabular}             & 2.080                        & 2.489                     \\
\hline
\begin{tabular}[c]{@{}l@{}}Out of Order\\ Packets(60\%)\end{tabular}             & 1.677                        & 23694.171                 \\
\hline
\begin{tabular}[c]{@{}l@{}}Low Bandwidth(1kbps\\ burst 256b)\end{tabular}        & 7.102                        & 10.704                    \\
\hline
\begin{tabular}[c]{@{}l@{}}Low Bandwidth (1kbps\\ burst 64b)\end{tabular}        & 9.869e+16                    & 9.272e+16                 \\
\hline
High Latency(200ms)                                                              & 1.984                        & 29.678                    \\
\hline
High latency(1200ms)                                                             & 1.440                        & 23693.894        \\
\hline
\end{tabular}}{}
\caption{\fontsize{10pt}{11pt}\selectfont{\itshape{Reconstruction loss with and without response time for each anomaly type}}}
\label{table:results}
\end{table*}

\subsection{Experiments on GAL-MAD}

We first conduct a comparative study with the state-of-the-art anomaly detection models on the newly introduced RS-Anomic dataset. The results in Table \ref{table:dataset_results_transposed} show that our model has achieved better performance across all the ratios. In particular, for all the ratios, the proposed GAL-MAD has achieved the highest recall compared to the existing anomaly detection methods, indicating the effectiveness of GAL-MAD in detecting anomalies. In anomaly detection, particularly within imbalanced datasets where anomalies are rare, recall (also known as sensitivity) is often the most critical metric. Recall measures the proportion of actual anomalies correctly identified by the model, ensuring that true anomalies are not overlooked. Missing an anomaly (a false negative) can have significant consequences.

Next, we evaluate the effect of including the response time as a feature in detecting the anomalies. Table \ref{table:results} shows average loss values for the two configurations, one using only cAdvisor metrics
and the other using both cAdvisor metrics and response time features. The response time features include moving averages calculated over 5-minute and 30-minute windows. The analysis reveals the significant impact of incorporating response time features on anomaly detection, particularly for network-related anomalies. Moreover, packet loss, out-of-order packets, and low bandwidth anomalies were tested under two different strengths to highlight the lack of sensitivity observed on fine network anomalies.

Finally, we conduct an ablation study to evaluate the impact of each component of the proposed GAL-MAD. Table \ref{table:AblationStudy} shows the results of the ablation study. The LSTM component was removed from the encoder and the decoder of the GAL-MAD detector to create the GAT-Autoencoder (GAT-AE), and both GAT layers were removed from the encoder and the decoder of the GAL-MAD detector to make the LSTM-Autoencoder (LSTM-AE). Finally, the GAT and the LSTM components were removed to construct a linear autoencoder (Linear-AE). The results demonstrate that the proposed GAL-MAD achieves the highest recall for all scenarios, indicating the importance of each component of the proposed architecture.

\begin{table*}[h!tb]
\centering
\setlength\tabcolsep{6pt}
\begin{tabular}{|l|l|l|l|l|l|}
\hline
\bfseries{Ratio} & \bfseries{Metric} & \bfseries{GAL-MAD} & \bfseries{GAT-AE} & \bfseries{LSTM-AE} & \bfseries{Linear-AE} \\
\hline
\multirow{3}{*}{95:5} 
& A & 0.9838 & 0.9808 & 0.9897 & \textbf{0.9893} \\
& R & \textbf{0.9884} & 0.8839 & 0.9032 & 0.8903 \\
& S & 0.9559 & 0.9867 & \textbf{0.9947} & 0.9953 \\
\hline
\multirow{3}{*}{90:10} 
& A & \textbf{0.9788} & 0.9704 & 0.9784 & 0.9784 \\
& R & \textbf{0.9228} & 0.8460 & 0.8492 & 0.8892 \\
& S & 0.9855 & 0.9867 & \textbf{0.9953} & \textbf{0.9953} \\
\hline
\multirow{3}{*}{60:40} 
& A & \textbf{0.9500} & 0.9264 & 0.9371 & 0.9358 \\
& R & \textbf{0.8938} & 0.8333 & 0.8462 & 0.8425 \\
& S & 0.9859 & 0.9857 & 0.9953 & \textbf{0.9955} \\
\hline
\end{tabular}
\caption{\fontsize{10pt}{11pt}\selectfont{\itshape{Ablation study of GAL-MAD Model in terms of Accuracy (A), Recall (R), and Specificity (S) for normal to anomaly ratios of 95:5, 90:10, and 60:40. AE stands for Auto Encoder.}}}
\label{table:AblationStudy}
\end{table*}

\subsection{Anomalous Service Localization Scores}

  \begin{figure}[!t]
  \centering
  \begin{subfigure}[b]{0.45\textwidth}
    \centering
    \includegraphics[width=\linewidth]{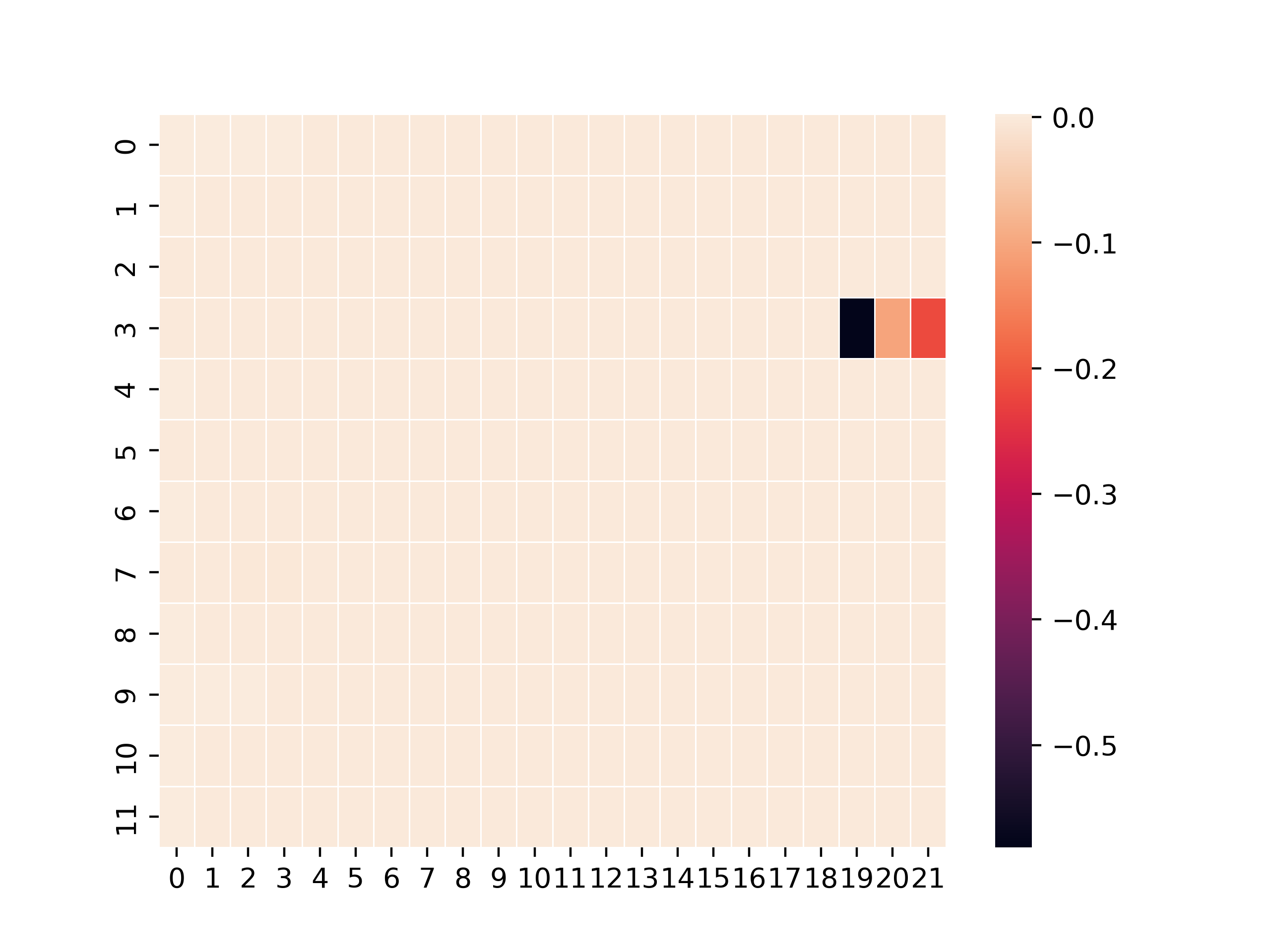}
    \caption{Response time delay anomaly in the MongoDB service}
    \label{heatmap_rt_delay_catalogue}
  \end{subfigure}
  \hfill
  \begin{subfigure}[b]{0.45\textwidth}
    \centering
    \includegraphics[width=\linewidth]{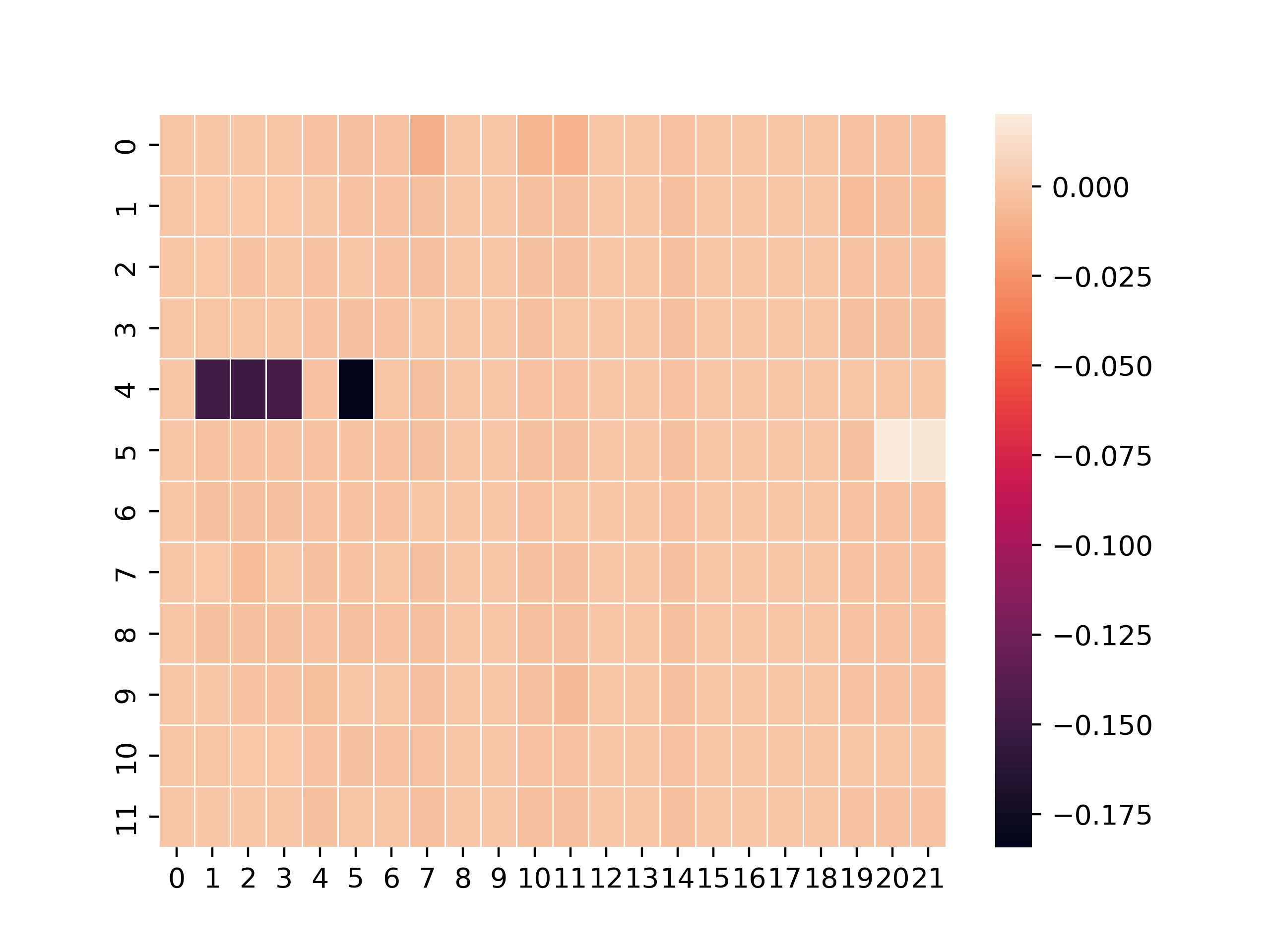}
    \caption{High CPU usage in the Dispatch service anomaly}
    \label{heatmap_cpu_dispatch}
  \end{subfigure}
  \vspace{0.25cm}
  \begin{subfigure}[b]{0.45\textwidth}
    \centering
    \includegraphics[width=\linewidth]{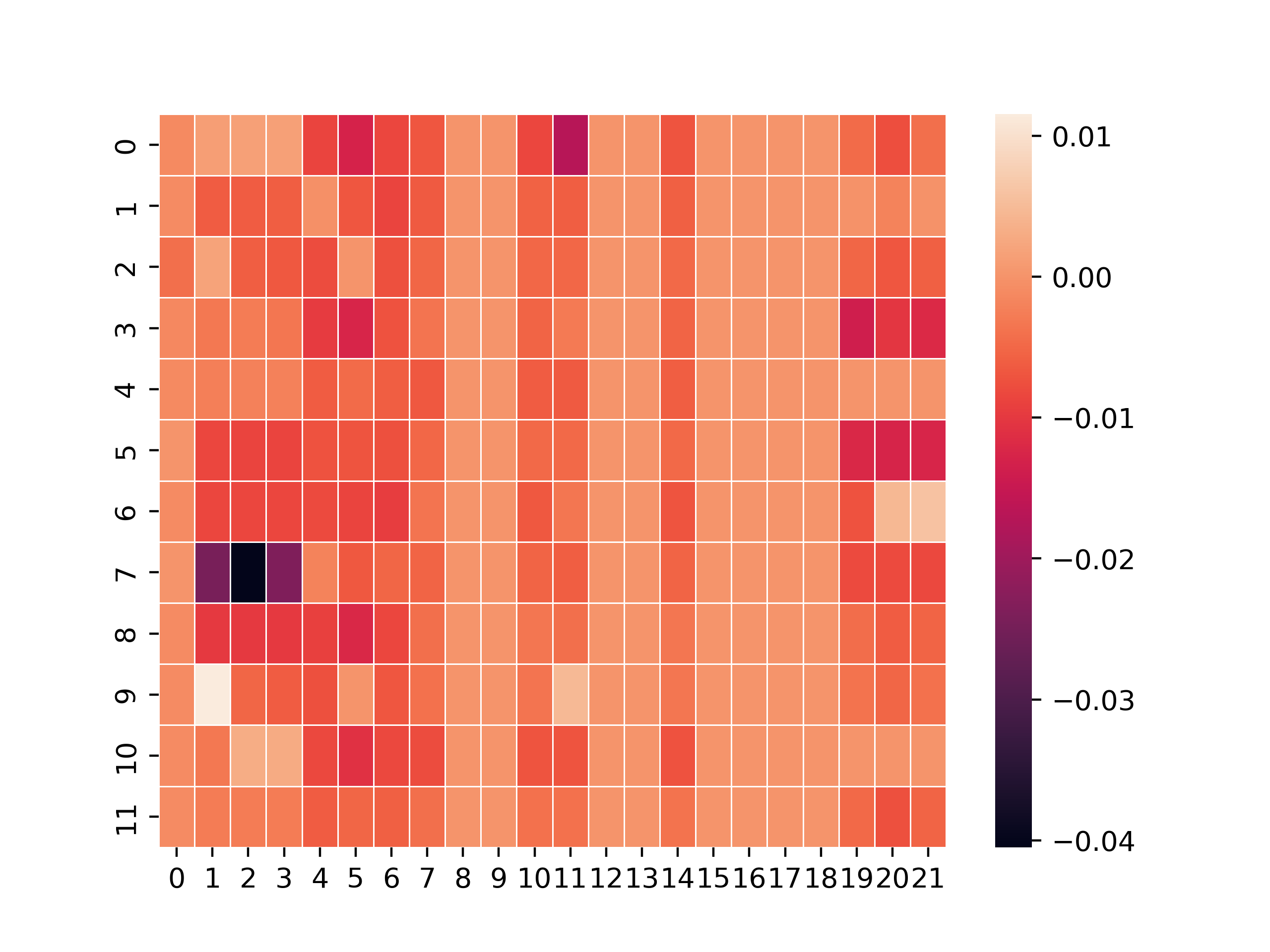}
    \caption{Out-of-order packets in the User service anomaly}
    \label{heatmap_out_of_order_packets}
  \end{subfigure}
  \hfill
  \begin{subfigure}[b]{0.45\textwidth}
    \centering
    \includegraphics[width=\linewidth]{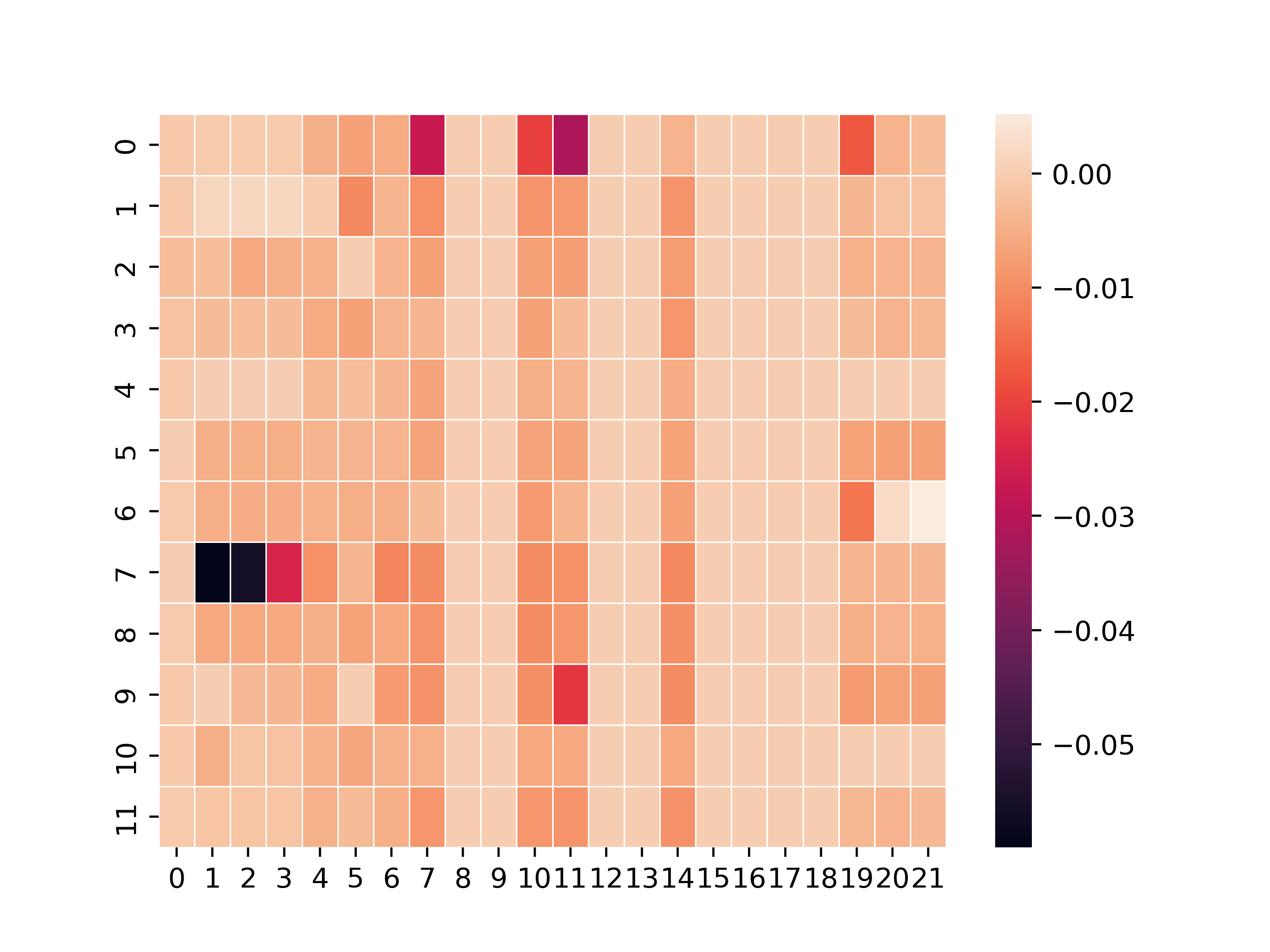}
    \caption{Packet loss anomaly in the User service}
    \label{heatmap_packet_loss_user}
  \end{subfigure}
  \caption{Heatmaps illustrating various anomalies detected in microservices}
  \label{fig:heatmaps_2x2}
\end{figure}

As described in Section \ref{anomaly_service_locatiozation}, we use eXplainable Artificial Intelligence (XAI) technique, SHAP \cite{shap}, to localize the anomalous service once an anomaly is detected. SHAP values were computed and visualized as heatmaps for selected anomalous instances, as depicted in Figures \ref{heatmap_rt_delay_catalogue} to \ref{heatmap_packet_loss_user}. In these heatmaps, the x-axis represents features (performance metrics) labelled from 0 to 21, corresponding to the sequence outlined in Table \ref{table:feature_details}. Specifically, labels 20 and 21 denote the moving averages of response times over 5-minute and 30-minute intervals, respectively. The y-axis enumerates the microservices in the following order: payment, shipping, redis, mongodb, dispatch, rabbitmq, user, mysql, catalog, ratings, web, and cart. 

The heatmaps offer intuitive insights, particularly for anomalies related to performance and response times. For instance, Figure \ref{heatmap_rt_delay_catalogue} illustrates divergent SHAP values for response time features when API calls from the catalog service to the mongodb service experienced delays. Similarly, Figure \ref{heatmap_cpu_dispatch} highlights significant SHAP values associated with CPU usage features in the dispatch service during instances of artificially induced high CPU utilization. Conversely, non-critical network-related anomalies, such as out-of-order packets and packet loss, pose challenges for visual interpretation. This is evident in Figures \ref{heatmap_out_of_order_packets} and \ref{heatmap_packet_loss_user}, where the SHAP value distributions are less distinct.

To enhance the detection of such anomalies, the SHAP values for the $container\_fs$ and response time features were amplified by a factor of four. The outcomes of this adjustment are documented in Table \ref{table:LocalizationResults}. In this table, an anomaly localization is deemed accurate when both the microservice experiencing the injected anomaly and its associated features are correctly identified.

\begin{table*}[h]
\centering
\setlength\tabcolsep{6pt}
\begin{tabular}{|l|l|l|}
\hline
\bfseries{Anomaly}                  & \bfseries{True service localization} & \bfseries{True feature localization} \\ \hline
rt-delay        & 10/10                        & 10/10                        \\ \hline
high-cpu         & 10/10                        & 10/10                        \\ \hline
high-fileIO       & 10/10                        & 10/10                        \\ \hline
memory-leak          & 10/10                        & 3/10                         \\ \hline
% service-down      & 1/10                         & 1/10                         \\ \hline
low-bandwidth        & 0                            & 0                            \\ \hline
out-of-order-packets & 0                            & 0                            \\ \hline
high-latency         & 6/10                         & 6/10                         \\ \hline
packetloss           & 9/10                         & 9/10                         \\ \hline
\end{tabular}
\caption{\fontsize{10pt}{11pt}\selectfont{\itshape{Anomalous service and feature localization results}}}
\label{table:LocalizationResults}
\end{table*}

\section{Conclusion}

This study introduces the RS-Anomic dataset, a comprehensive multivariate time-series resource designed to advance anomaly detection research in microservices architectures. The dataset comprises 100,000 normal and 14,000 anomalous data points, encompassing ten distinct anomaly types, providing a robust foundation for academic and industrial research endeavors. Building upon this dataset, we propose the Graph Attention and LSTM-based Microservice Anomaly Detection (GAL-MAD) model, which leverages Graph Attention Networks and Long Short-Term Memory networks to capture spatial and temporal dependencies among microservices. Empirical evaluations demonstrate that GAL-MAD outperforms existing state-of-the-art anomaly detection models on the RS-Anomic dataset, achieving superior recall across varying anomaly ratios. To enhance interpretability, SHAP  values were employed, facilitating the localization of anomalous services and identifying contributing features, particularly in performance-related anomalies. This explainability empowers system administrators to diagnose anomalies swiftly, comprehend the root causes of performance degradations, and gain deeper insights into microservice interactions.

The RS-Anomic dataset and the GAL-MAD model represent significant strides toward intelligent and interpretable anomaly detection in complex distributed systems. Future research directions include expanding the dataset to incorporate multi-node architectures and auto-scaling capabilities that more accurately reflect production-level microservices deployments. Additionally, while SHAP provides valuable insights into root cause localization, further work is warranted to interpret anomalies within the context of application architecture. Moreover, in-depth analysis of network-related metrics is essential to enhance the explainability and diagnostic precision of network anomalies.

% \begin{figure}[H]
%   \centering
%       \includegraphics[width=0.75\textwidth]{Figures/heatmaps/rt_delay_catalogue.png}
%       \caption{Heat map for response time delay anomaly in the mongo database service. Service 3 represents the MongoDB service. Features 19, 20, and 21 represent response times and moving averages of the response times for 5 and 30 minute windows, respectively.}
%      \label{heatmap_rt_delay_catalogue}
%   \end{figure}

% \begin{figure}[H]
%   \centering
%       \includegraphics[width=0.75\textwidth]{Figures/heatmaps/cpu_dispatch.png}
%      \caption{Heat map for high CPU usage in the dispatch service anomaly. Service 4 represents the dispatch service. Features 1,2,3, and 5 represent memory usage, memory failures, memory working set usage and CPU usage time, respectively.}
%      \label{heatmap_cpu_dispatch}
%   \end{figure}
%   \begin{figure}[H]
%   \centering
%       \includegraphics[width=0.75\textwidth]{Figures/heatmaps/out_of_order_packets.png}
%        \caption{Heat map for a false localization of out-of-order packets in the user service.}
%      \label{heatmap_out_of_order_packets}
%   \end{figure}
%   \begin{figure}[H]
%   \centering
%       \includegraphics[width=0.75\textwidth]{Figures/heatmaps/packet_loss_user.png}
%        \caption{Heat map for a false localization of packet loss anomaly in the user service.}
%      \label{heatmap_packet_loss_user}
%   \end{figure}

\bibliographystyle{unsrtnat}
\bibliography{arxiv} 

\end{document}